\def \SAIT #1 #2 {{\em Mem.\ Soc.\ Astron.\ It.\/} {\bf #1}, #2}
\def \MESS #1 #2 {{\em The Messenger\/} {\bf #1}, #2}
\def \ASTRNACH #1 #2 {{\em Astron. Nach.\/} {\bf #1}, #2}
\def \AAP #1 #2 {{\em Astron. Astrophys.\/} {\bf #1}, #2}
\def \AAL #1 #2 {{\em Astron. Astrophys. Lett.\/} {\bf #1}, L#2}
\def \AAR #1 #2 {{\em Astron. Astrophys. Rev.\/} {\bf #1}, #2}
\def \AAS #1 #2 {{\em Astron. Astrophys. Suppl. Ser.\/} {\bf #1}, #2}
\def \AJ #1 #2 {{\em Astron. J.\/} {\bf #1}, #2}
\def \ANNREV #1 #2 {{\em Ann. Rev. Astron. Astrophys.\/} {\bf #1}, #2}
\def \APJ #1 #2 {{\em Astrophys. J.\/} {\bf #1}, #2}
\def \APJL #1 #2 {{\em Astrophys. J. Lett.\/} {\bf #1}, L#2}
\def \APJS #1 #2 {{\em Astrophys. J. Suppl.\/} {\bf #1}, #2}
\def \APSS #1 #2 {{\em Astrophys. Space Sci.\/} {\bf #1}, #2}
\def \ASR #1 #2 {{\em Adv. Space Res.\/} {\bf #1}, #2}
\def \BAIC #1 #2 {{\em Bull. Astron. Inst. Czechosl.\/} {\bf #1}, #2}
\def \JSQRT #1 #2 {{\em J. Quant. Spectrosc. Radiat. Transfer\/} {\bf #1}, #2}
\def \MN #1 #2 {{\em Mon. Not. R. Astr. Soc.\/} {\bf #1}, #2}
\def \MEM #1 #2 {{\em Mem. R. Astr. Soc.\/} {\bf #1}, #2}
\def \PLR #1 #2 {{\em Phys. Lett. Rev.\/} {\bf #1}, #2}
\def \PASJ #1 #2 {{\em Publ. Astron. Soc. Japan\/} {\bf #1}, #2}
\def \PASP #1 #2 {{\em Publ. Astr. Soc. Pacific\/} {\bf #1}, #2}
\def \NAT #1 #2 {{\em Nature\/} {\bf #1}, #2}

\documentstyle{memsait}
\input epsf.sty
\begin{opening}
\title{THE COLOURS OF THE X--RAY BACKGROUND} 
\author{ANDREA COMASTRI}
\institute{Osservatorio Astronomico di Bologna, via Zamboni 33, 40126 
Bologna, Italy}
\date{} 
\end{opening}

\begin{document}

\oddpagefooter{}{}{} 
\evenpagefooter{}{}{} 
\ 
\bigskip

\begin{abstract}

The recent deep X--ray surveys at both soft (0.5--2.0 keV) and hard
(2--10 keV) energies have greatly extended our knowledge of the 
X--ray source density and spectral shapes at relatively faint
fluxes adding further evidence on the fact that discrete sources, 
mainly AGNs, are responsible for the X--ray background (XRB) emission
over a broad energy range.
In addition the first complete optically identified samples 
of soft X--ray sources are becoming available allowing 
to test the XRB AGN synthesis models in the light of recent results. 
In this paper I will briefly compare the model predictions with some
new observational data. 

\end{abstract}

\section{AGN models for the XRB}

It has been recognized already ten years ago (Setti \& Woltjer 1989) that 
the XRB spectral paradox  (i.e. the fact that the source spectra in the 2--10 
keV energy range are much steeper than the XRB spectrum) can be solved 
assuming an important contribution from sources with spectral shapes 
flattened by absorption. 

In the framework of popular AGN unified schemes
the 3--100 keV XRB spectrum and the source counts in different energy bands
can readily be reproduced by the combined emission of Seyfert galaxies
and quasars (type 1) and obscured (type 2) AGNs with a range of column 
densities and luminosities
(Matt \& Fabian 1994; Madau et al. 1994; Comastri et al. 1995 hereinafter C95). 
  
All these models rely on several assumptions on the sources spectral shapes,
X--ray luminosity function (XLF) and cosmological evolution.
Given the large number of ``free parameters" is not surprising that the
XRB spectrum and intensity can be well reproduced even for rather different
choiches of the above described parameters. 
An attempt to reduce the parameter space has been made by 
C95 who developed a self-consistent approach 
which simultaneously fit the XRB spectrum and others available 
constraints: namely the source counts, redshift and absorption distributions 
observed in the soft (0.5--2 keV) and hard (2--10 keV) bands.
Assuming for the spectral and evolutive AGN properties
average values consistent with the observations 
the absorption distribution of type 2 objects is the only ``free" parameter 
which is varied until a global good fit is obtained (see Fig. 1 and figures
in C95). According to this model the XRB spectrum is due 
to sources with a similar range of redshifts and
luminosities, but with very different spectral shapes (colours).
The optical identifications of faint X--ray sources discovered by
ROSAT, ASCA and BeppoSAX surveys open the possibility to 
test the AGN synthesis model for the XRB. Two examples are briefly discussed 
in the following.

\begin{figure}
\epsfysize=9cm 
\hspace{-6cm}\epsfbox{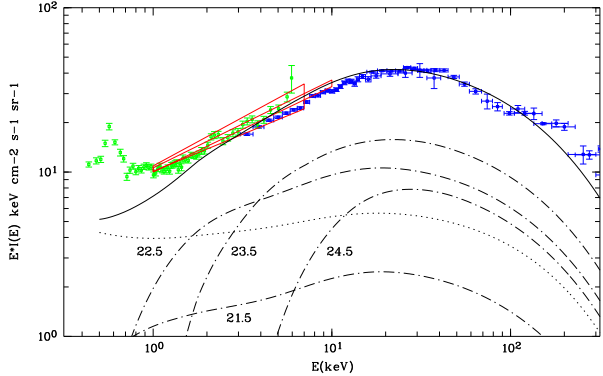} 
\caption[h]{The AGN synthesis model (solid line) fit to the XRB spectrum. 
The various curves represent the contribution of unabsorbed (dotted line) 
and absorbed (dot--dashed lines) objects. There is a clear mismatch between
the 0.4--6 keV ASCA data (grey points), the 1--7 keV ASCA plus ROSAT joint fits
(dotted bowties) and the HEAO1--A2 data above 3 keV (black points) in the 
overlapping energy range. The origin of 
this difference, which is of the order of 20--30 \% in the 1--7 keV range, is
not yet understood.}
\end{figure}
\begin{figure}
\hspace{-6cm}
\epsfxsize=5.0cm
\epsfbox{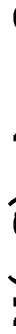}
\hspace{2.2cm}
\epsfxsize=5.0cm
\epsfbox{comastri_fig3.ps}
\caption[h]{{\it Left panel}: the 0.5--2 keV counts compared with model 
predictions (line types as in Fig. 1). {\it Right panel}: the colour-colour  
diagram of the optically identified RDS sources (see text for details). Data
points for both panels are from Hasinger et al. 1998}
\end{figure}

\section{The AGN content of deep X--ray surveys}

The contribution to the XRB of the absorbed objects increases 
with energy (Fig. 1), as a consequence a significant test for this model 
would be the comparison of its predictions with the results of optical 
identifications of complete samples of X--ray sources possibly selected in the 
hard X--ray band. 
Programs to optically identify these sources have 
already started, but the number of identified sources is still low and a 
detailed comparison will be possible only in the near future.
In order to overcome this problem I have considered the complete 
optically identified sample of the Rosat Deep Survey (RDS) in the 
Lockman hole (Hasinger et al. 1998). 
Among the 50 X--ray sources with 0.5-2 keV fluxes $>$ 5.5 10$^{-15}$ erg 
cm$^{-2}$ s$^{-1}$ the great majority (43) are AGNs (Schmidt et al. 1998). 
At this limiting flux the model predicts (Fig. 2, left panel) a fraction of 
absorbed
AGN ($N_H=10^{21-23}$) of the order of 30 \% indicating that the bright 
tail of the absorbed population should be visible in the RDS.
X--ray spectral information for the RDS sources is only available in terms 
of the two hardness ratios HR1 and HR2 (see Hasinger et al. 
1998 for the definition) and is reported in Figure 2. 
Absorbed sources are characterized by hard X--ray spectra and populate the 
upper right and right parts of the diagram. The hardenss ratios of obscured AGNs
with a range of column densities ($N_H=10^{21-23}$) and redshifts
($z=0-3$) assuming a power law spectrum ($\alpha=1$) lie on the 
continuous lines.
At this limiting flux about 10--12 sources have hardness ratios and redshifts 
which are consistent with the presence of such column densities, moreover 
the optical spectra of most of these hard sources show only
narrow lines which are typical of type 2 AGNs. 
Even if the agreement with the model predictions is rather good 
the number of absorbed sources is still too low preventing from a more 
detailed analysis. It should be noted that the model predictions, based on the
latest response matrix, 
do not reproduce the HR2 distribution of unabsorbed sources.
This discrepancy may be due to either a much steeper, than assumed, 
0.5--2.0 keV spectral index, or to residual calibration 
uncertainities. Given that the fraction of absorbed sources increases at 
lower fluxes (Fig. 2; left panel) the optical
identification of faint sources coupled with the X--ray spectral information  
is expected to provide further constraints on the model parameters.
In particular the comparison of the expected and observed luminosity/redshift 
distributions of the absorbed sources would provide a key test for the model.
 
\section{Do quasar 2 exist ?}

The existence of a population of highly absorbed intrinsically
luminous sources, which in analogy with the Seyfert 1/2
classification are named quasars 2, is predicted by the AGN unification schemes 
at least in their strictest version. Based on these schemes the C95 
model assumes the same XLF and pure luminosity evolution for type 1 and type 2 
(obscured) objects, as a consequence several type 2 QSOs are expected in 
medium--deep X--ray surveys. Even if the discovery of a few type 2 QSO 
candidates has been reported (e.g. Otha et al. 1996) this population has 
proved to be elusive 
and its existence has been recently questioned (Halpern et al. 1998).
The surface density (deg$^{-2}$) of type 2 QSOs according to C95 is 
reported in Table 1 for two different limiting fluxes (cgs units) and 
energy ranges (keV units) corresponding to the limits of ASCA and
ROSAT deep surveys.
As the definition of a type 2 QSO is somewhat ambiguous the 
calculations have been performed for different choiches of the 
minimum column density and luminosity (log values) required to classify a 
source as a type 2 QSO. 
Taking into account the sky coverage of the RDS the number of type 2 
quasar ranges from 0.1 to 4. There are 3 RDS sources with hardness ratios
and luminosities which meet the above criteria.
The comparison with the observations in the 2--10 keV band is not 
straigthforward given the large incompletness of the optical identifications.
Among the 40 X--ray sources in the ASCA Large Sky Survey (Ueda et al. 1998) 
the expected number of type 2 QSOs (according to Table 1) ranges from 
1 to about 10.  

\vspace{0.5cm} 
\centerline{\bf Tab. 1 - The space density of quasar 2}
\begin{table}[h]
\hspace{0.1cm} 
\begin{tabular}{|l|c|c|c|c|}
\hline
  & N$_H>$22, L$>$44 & N$_H>$22, L$>$45 & N$_H>$23, L$>$44 & N$_H>$23, L$>$45 \\
\hline
S$_{\rm 2-10} > 10^{-13}$ & 4.2  & 0.9   & 1.5 & 0.5 \\
S$_{\rm 0.5-2} > 5 \cdot 10^{-15}$ & 39 & 17  & 0.7 & 0.7 \\
\hline
\end{tabular}
\end{table}

It is concluded that the existence of type 2 QSOs cannot be 
ruled out on the basis of the present observations. Given that 
their surface density increases steeply a large number of quasars 2 is 
expected at faint X--ray fluxes (Comastri 1999 in preparation).
Alternatively a large population of lower luminosity Seyfert 2 galaxies, 
possibly subject to a different cosmological evolution law, may account for 
the bulk of the XRB at high energies. It is interesting to note 
that some evidence of a more complex beahviour than pure luminosity evolution
is emerging from ROSAT data (Hasinger 1998). If this is the case
a revision of the model assumptions will be necessary.
Deep observations with AXAF and XMM will clarify some of these issues.

\end{document}